\documentclass[aps,prd,10pt,notitlepage,nofootinbib]{revtex4-1}

\usepackage[utf8]{inputenc}
\usepackage{amsmath,amssymb,amsfonts}
\usepackage{graphicx, overpic}
 \usepackage{palatino}

\usepackage{bbold}
\usepackage{bm}
\usepackage[usenames,dvipsnames]{xcolor}
\usepackage{color}
\usepackage[colorlinks=true,linkcolor=blue,urlcolor=blue,citecolor=blue]{hyperref}

\usepackage{slashed}
\usepackage[english]{babel}
\usepackage{pifont}
\usepackage{dsfont,mathrsfs}
\usepackage{cancel}
\usepackage{bigints}
\usepackage{accents}
\usepackage{soul}
\usepackage{multirow}
\usepackage{natbib}
\usepackage{tikz-feynman}

\usepackage{amsmath, amssymb}
\usepackage{bbold}
\usepackage{makecell}
\usepackage{simpler-wick}
\usepackage[export]{adjustbox}

\usepackage[bb=boondox]{mathalpha}

\usepackage{tikz}
\usepackage{tikz-feynman}
\tikzfeynmanset{compat=1.1.0} 

\usetikzlibrary{decorations.pathmorphing, decorations.markings, arrows.meta}

\usepackage{orcidlink} 

\newcommand{\nn}{\nonumber}

\newcommand{\MB}[1]{\left|#1\right|}
\newcommand{\FB}[1]{\left(#1\right)}

\newcommand{\TB}[1]{\left[#1\right]}

\newcommand{\IM}{\text{Im}}

\newcommand{\half}{\dfrac{1}{2}}
\newcommand{\gm}{\gamma}

\newcommand{\ov}[1]{\overline{#1}}

\newcommand{\mM}{\mathcal{M}}

\newcommand{\Li}{{\rm Li}_2 }
\newcommand{\pLi}{{\rm Li}_2^\prime  }

\allowdisplaybreaks

\begin{document}
	\title{Thermal photon production rate from QGP at finite density: Analytic cutoff independence}

	\author{Sourav Duari\orcidlink{0009-0006-0795-5186}$^{a,c}$}
	\email{s.duari@vecc.gov.in}
	\email{sduari.vecc@gmail.com}
	
	\author{Nilanjan Chaudhuri\orcidlink{0000-0002-7776-3503}$^{a}$}
	\email{n.chaudhri@vecc.gov.in}
	\email{nilanjan.vecc@gmail.com}

\author{Sourav Sarkar\orcidlink{0000-0002-2952-3767}$^{a,c}$}
\email{sourav.vecc@gmail.com}
	
	\author{Pradip Roy\orcidlink{0009-0002-7233-4408}$^{b,c}$}
	\email{pradipk.roy@saha.ac.in}

	\affiliation{$^a$Variable Energy Cyclotron Centre, 1/AF Bidhannagar, Kolkata - 700064, India}
	\affiliation{$^b$Saha Institute of Nuclear Physics, 1/AF Bidhannagar, Kolkata - 700064, India}
	\affiliation{$^c$Homi Bhabha National Institute, Training School Complex, Anushaktinagar, Mumbai - 400085, India}
%
	
	\begin{abstract}
        We compute the thermal photon production rate from a hot and dense quark-gluon plasma. While the mutual cancellation of the intermediate momentum cutoff $k_c$ between the hard and soft regimes in the presence of finite baryon density has been previously established only numerically, we demonstrate here, for the first time, that this cutoff independence can be shown analytically even in the presence of a nonzero quark chemical potential $\mu$. This resolves a long-standing issue addressed in~\cite{Traxler:1994hy}. The final expression for the photon rate can be treated as a generalization of the results obtained in~\cite{Kapusta:1991qp,Baier:1991em} in the presence of $\mu$ and can be implemented directly in the space-time evolution of a hot and dense QGP medium using relativistic hydrodynamics.
    \end{abstract}
	
	\maketitle
		\section{Introduction}
\label{Sec_Intro}
 
 Photons have long been considered to be an effective probe of relativistic heavy-ion collisions.In this work we focus on the thermal photon production rate from a quark-gluon plasma (QGP) in the presence of a finite chemical potential. This is essential as heavy ion collision experiments at the Relativistic Heavy Ion Collider (RHIC) for example, can typically create a medium where the net baryon density is non-vanishing~\cite{Andronic:2005yp,Braun-Munzinger:2011fyo,Andronic:2017pug}. Moreover, recent beam-energy–scan programs at RHIC and future experiments at FAIR and NICA indicate a renewed interest in understanding the dynamics of QGP at finite density~\cite{CBM:2016kpk,STAR:2017sal,Senger:2021cfo}. As we enter a new era of precision measurements, it is therefore important to consider the effect of a non-vanishing baryon chemical potential on the thermal photon yield from the QGP. The presence of finite chemical potential modifies the thermal distribution functions of quarks and antiquarks associated with the photon production processes, eventually leading to an enhanced production rate from the hot and dense QGP medium. 
 
The contribution to the hard photon production from quark-antiquark annihilation and QCD Compton scattering is computed to leading order in perturbation theory~\cite{Kapusta:1991qp,Baier:1991em,Traxler:1994hy,Duari:2026ejq}. In order to regulate the infrared singularities  of the invariant amplitude a momentum cutoff $k_c$ is introduced which separates the hard and soft momentum regions. The cancellation of this intermediate scale in the total rate at finite temperature and density is a crucial issue. At finite quark chemical potential this cancellation has been verified only numerically in Refs.~\cite{Traxler:1994hy,Duari:2026ejq}. Here, we demonstrate this cancellation analytically for the first time in the presence of a nonzero quark chemical potential $\mu$, thereby resolving this long-standing issue. Thus we provide an extension of the framework developed in~\cite{Kapusta:1991qp,Baier:1991em} in presence of finite baryon density.

\section{Setup}

The real photon emission rate with energy $E$ and momentum $\vec{p}$ from a thermalized  plasma of quarks, antiquarks and gluons is given by~\cite{Kapusta:1991qp,Baier:1991em,Kapusta:2006pm,Bellac:2011kqa,Haque:2024gva}
\begin{equation}
	E \frac{dR}{d^3 p} = \frac{2}{(2\pi)^3} ~\IM~ ^{\rm ret}\Pi^\mu_\mu (E,\vec{p}) ~\frac{1}{e^{\beta E} -1} \label{Eq_Rate}
\end{equation}
where $^{\rm ret}\Pi^\mu_\mu$ is the trace of the in-medium  (retarded) photon self-energy. This expression is exact to all orders in the strong coupling, but only to leading order in the electromagnetic coupling, since it relies on the assumption that the emitted photon escapes the thermal medium without undergoing further final-state interactions~\cite{Kapusta:1991qp,Kapusta:2006pm}. 
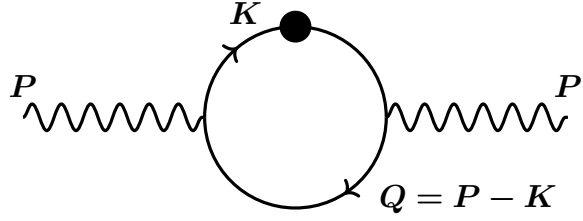
\begin{figure}[h]
\begin{tikzpicture}[scale=1.2]
	
	\def\R{1.}
	\coordinate (Lv) at (-\R,0);   
	\coordinate (Rv) at (\R,0);    
	
	\draw[line width=1.2pt, decorate,
	decoration={snake, amplitude=5 pt, segment length=11pt}
	] (-3.,0.) -- (Lv);
	
	\draw[line width=1.2pt, decorate,
	decoration={snake, amplitude=5pt, segment length=11pt}
	] (Rv) -- (3.,0);
	
	\node at (-3.,0.35) {\large $\bm{P}$};
	\node at (3.,0.35) {\large $\bm{P}$};
	
	\draw[line width=1.2pt,
	postaction={decorate},
	decoration={markings, mark=at position 0.28 with {\arrow{>}}}
	] (Lv) arc[start angle=180, end angle=0, radius=\R];
	
	\draw[line width=1.2pt,
	postaction={decorate},
	decoration={markings, mark=at position 0.72 with {\arrow{<}}}
	] (Lv) arc[start angle=180, end angle=360, radius=\R];
	
	\node[circle, fill=black, minimum size=12pt, inner sep=0pt] (blob) at (0,\R) {};
	
	\node at (-0.55,\R+0.15) {\large $\bm{K}$};
	
	\node at (1.9,-\R+0.1) {\large $\bm{Q=P-K}$};
	
\end{tikzpicture}
	\caption{Photon self-energy computed in the HTL approximation. The internal fermion line marked with a black blob denotes the soft quark propagator resummed within the HTL scheme, while the unmarked line represents the bare hard quark propagator.} 
	\label{Fig_photon}
\end{figure}
It is well established that tree-level amplitudes involving the exchange of massless particles exhibit infrared singularities, which get regulated once thermal loop effects are incorporated~\cite{Braaten:1991dd}. Accordingly, the photon emission rate can be separated into hard and soft pieces, classified by the magnitude of the exchanged quark momentum. For computing the hard contribution, for which the momentum of the exchanged quark is large, it is sufficient to use bare propagators within the standard perturbative approach. However, in the soft-momentum regime the exchanged quark must be described using a thermally resummed (HTL) propagator. Following the weak-coupling framework, one introduces an intermediate momentum scale $k_c$ which satisfies $ gT \ll k_c \ll T$. Here $g$ is the strong coupling constant and  $T$ represents the medium temperature. This cutoff parameter $k_c$ serves as a demarcation between  the hard and soft sectors. Although the individual hard and soft contributions depend on this arbitrary scale $k_c$, this dependence cancels once the two pieces are combined to obtain the total photon production rate~\cite{Kapusta:1991qp,Baier:1991em,Kapusta:2006pm,Bellac:2011kqa}.

\subsection{Soft Contribution}
\label{Sec_softpart}

The soft part of the real photon emission rate from QGP at finite quark chemical potential has been obtained in~\cite{Traxler:1994hy,Haque:2024gva,Chaudhuri:2025kpl} from the imaginary part of the retarded photon self-energy as shown in Fig.~\ref{Fig_photon}. Since the photon energy is large, kinematics restricts the polarization tensor to contain only one soft quark propagator, while the other quark line remains hard. No vertex correction is required in this kinematic limit, since the hard photon couples to the soft quark line through the bare electromagnetic vertex. The analytical expression for the soft contribution to the thermal photon production rate with energy $E$ is given by
\begin{equation}
	2E\frac{dR^{\rm \ soft}}{d^3 p} = \dfrac{5 \alpha }{6 \pi^3} e^{-\beta E} { M^2 \ln \dfrac{k_c^2}{2 M^2} }~\label{Eq_Soft_cont}~.
\end{equation}
Here the thermal quark mass is defined as
\begin{align}
	&M^2 = \dfrac{g^2 C_F}{4 \pi^2} \int k ~dk \TB{ f_{FD}^{+} (k) + f_{FD}^{-} (k)+ 2 f_{BE} (k) } \nn \\
    &= - \dfrac{g^2 T^2 C_F}{4 \pi^2}\TB{ \Li\FB{-e^{\frac{\mu}{T}}} + \Li\FB{-e^{-\frac{\mu}{T}}} + 4 \Li\FB{-1}} \nn \\
	&= -\dfrac{g^2 T^2 C_F}{4 \pi^2} \TB{ -\dfrac{\pi^2}{6} - \half \dfrac{\mu^2}{T^2} - 4 \dfrac{\pi^2}{12}} \nn \\
	&= \dfrac{\pi \alpha_S C_F}{2} \FB{T^2 + \dfrac{\mu^2}{\pi^2}} \label{Eq_ML}~,
\end{align}
where $\Li (x) $ denotes the polylogarithm of order 2 and the distribution functions are given by 
\begin{align}
	f^{\pm}_{FD} (E) &= \dfrac{1}{\exp\FB{\dfrac{E\mp \mu}{T}}  +1}~, \\
    f_{BE} (E) &=\dfrac{1}{\exp\FB{\dfrac{E}{T}}  -1}~\label{Eq_dist_fns}~,
\end{align}
where $\mu$ is the quark chemical potential and $C_F = {4}/{3}$ is the Casimir invariant of the gauge group $SU(3)$ in the fundamental representation. Fine structure constant and strong coupling constant are defined as $\alpha = e^2/4\pi$ and $\alpha_S = g^2/4\pi$ respectively. The effective quark propagator at finite density, required for evaluating the soft contribution, has been derived in~\cite{Kapusta:2006pm,Bellac:2011kqa,Mustafa:2022got,Duari:2025iah}.  
While calculating the soft contribution as expressed by Eq.~\eqref{Eq_Soft_cont} we have assumed that the exchanged quark propagator is dressed and satisfies a hyperbolic cutoff $0 \le k^2 - \omega^2 \le k_c^2$ in accordance with Refs.~\cite{Kapusta:1991qp,Baier:1991em}.  The same cutoff is introduced in the hard contribution to regulate the infrared divergences associated with the soft region of phase space in the $t$- and $u$-channel diagrams. In the limit $\mu\to0$, Eq.~\eqref{Eq_Soft_cont} reduces to the well-known result of Refs.~\cite{Kapusta:1991qp,Baier:1991em}.

\subsection{Hard Contribution}
\label{Sec_hardpart}

The hard contribution to the real photon production rate is obtained from the scattering matrix elements for quark-antiquark annihilation and QCD Compton processes to leading order in perturbation theory. These have been worked out in detail in Refs.~\cite{Kapusta:1991qp,Baier:1991em,Traxler:1994hy,Duari:2026ejq}, and only the final results are quoted below. For the pair-annihilation process, the squared amplitude is given by
\begin{align}
	\sum \MB{\mM^{q\bar{q}\to g \gm }_{\rm tot}}^2 =  \dfrac{5\pi^2}{9}2^9 \alpha \alpha_S \FB{ \dfrac{t}{u} +\dfrac{u}{t}   }\label{Eq_M2_anni}~.
\end{align}
Here $\sum$ represents sum over spins, colors and flavors. For QCD Compton processes for quarks and gluon the squared amplitude is expressed as
 \begin{equation}
 	\sum \MB{\mM^{qg\to q\gm}_{\rm tot}}^2 =  \dfrac{5\pi^2}{9}2^9 \alpha \alpha_S \FB{ \dfrac{t}{s} +\dfrac{s}{t}   }~.  \label{Eq_M2_comp}
 \end{equation}
 The corresponding antiquark Compton process has the same squared amplitude by charge-conjugation invariance. Here, $s$, $t$, and $u$ denote the Mandelstam variables.
 Since we have the squared amplitudes for all three possible processes (annihilation and two Compton processes for quark and antiquark), we can now obtain the hard part of the photon production rate. All strongly interacting particles participating in these processes are assumed to be in thermal equilibrium, whereas the emitted photon is not re-scattered because of its large mean free path. These processes contribute to the differential rate which is given by~\cite{Kapusta:1991qp,Baier:1991em} 
\begin{widetext}
\begin{align}
	2E\frac{dR^{\rm \ hard}}{d^3 p} = \dfrac{1}{(2\pi)^8} \int \dfrac{d^3 \vec{p}_1}{2 E_1} \dfrac{d^3 \vec{p}_2}{2 E_2} \dfrac{d^3 \vec{p}_3}{2 E_3} \delta^{(4)} \FB{P_1 + P_2 -P_3 - P} f_1(E_1)f_2(E_2) \FB{1\pm f_3(E_3)} \sum\MB{\mM_{\rm tot}^{ \rm process}}^2 \label{Eq_rate_hard1} 
\end{align}
Here, \(f_i\) denotes the equilibrium distribution function of the corresponding incoming or outgoing strongly interacting particle. The factor \(1\pm f_3(E_3)\) accounts for Bose enhancement \((+)\) for the final-state gluon in annihilation and Pauli blocking \((-)\) for the final-state quark or antiquark in the Compton processes. Assuming the outgoing photon is along $z$-direction, nine integrations appearing in Eq.~\eqref{Eq_rate_hard1} can be reduced to a four-dimensional form~\cite{Haque:2024gva,Duari:2026ejq}. This yields
\begin{align}
		2E\frac{dR^{\rm \ hard}}{d^3 p} = \dfrac{1}{8 E (2\pi)^7} \int_{2 k_c^2}^{\infty}ds \int_{-s +k_c^2}^{-k_c^2} dt  \sum\MB{\mM_{\rm tot}^{ \rm process}}^2 \int dE_1 dE_2 \dfrac{\Theta\FB{\Xi (E_1,E_2)}  f_1 f_2 \FB{1\pm f_3} }{\sqrt{\Xi\FB{E_1,E_2}}} \label{Eq_rate_hard2}
\end{align}
where
\begin{align}
	\Xi \FB{E_1,E_2} &= - (s+t)^2 E_2^2 + 2 (s+t) (sE-t E1)+ st (s+t) - (sE + t E_1)^2
\end{align}
\end{widetext}
and $\Theta $ is the step function. As mentioned earlier, $k_c$ is an infrared cutoff introduced to tame the singularities of the invariant amplitudes as $t$ and/or $u$ goes to zero. For hard photon production, the relevant kinematic regime is \(E_1+E_2>E\gg T\). In this regime, the incoming distribution functions $f_1$ and $f_2$ can be approximated by Boltzmann distribution functions instead of Fermi-Dirac or Bose-Einstein  forms, such that $f_1 (E_1)f_2 (E_2) \approx e^{-(E_1+E_2)/T}$. This allows one to perform all the integrals analytically at $\mu = 0$ which leads to a closed-form expression for the differential rate of hard photon production from QCD Compton and annihilation processes~\cite{Kapusta:1991qp,Baier:1991em,Haque:2024gva}. At finite $\mu$, a similar approach has been previously found to yield a hard part that does not correctly reproduce the $\mu\to 0$ limit. This is because  the resulting closed-form expression, as discussed in Ref.~\cite{Dumitru:1993us}, contains a term $\sim {\rm Ei}\FB{(4 \mu E - k_c^2)/4 E T}$ which results in a branch point at finite $\mu$. This difficulty has since prevented a fully analytic demonstration of the cutoff independence of the total photon emission rate at finite density obtained by summing the soft and hard contributions. In Ref.~\cite{Traxler:1994hy}, the hard contribution was instead evaluated numerically using Eq.~\eqref{Eq_rate_hard2}. Combining this with the analytic soft contribution given by Eq.~\eqref{Eq_Soft_cont}, it has been demonstrated numerically that the total photon rate is independent of the arbitrary separation parameter $k_c$.

However, on closer examination, we find that the derivation of Ref.~\cite{Dumitru:1993us} is inconsistent. In fact, the hard and soft contributions can be combined analytically in a $k_c$-independent closed form, even in the presence of a finite quark chemical potential $\mu$. In the following sections, we demonstrate analytically that the hard contribution can be evaluated consistently at finite \(\mu\), allowing its cutoff dependence to cancel exactly against that of the soft contribution, resolving the long-standing issue identified in Refs.~\cite{Dumitru:1993us,Traxler:1994hy}.

\section{Simplification of the hard part for high energy photons}
\label{Sec_cutoff}

 To simply Eq.~\eqref{Eq_rate_hard2} we follow the procedure discussed in~\cite{Duari:2026ejq}. Let us introduce $V = E_1 + E_2$, then replacing $E_1$ by $E_1 = V - E_2$, we can write 
\[
\Xi(E_2, V) = \left( A E_2^2 + B E_2 + C \right)^{1/2}
\]
where
\begin{align}
	A &= -s^2 \\
	B &= 2s \left( sE + 2tE - tV \right) \\
	C &= st (s + t) - (sE + tV)^2~.
\end{align}
Limits of these integrations over $E_2$ and $V$ are fixed by kinematical constraints  appearing from the energy-momentum conserving delta function in Eq.~\eqref{Eq_rate_hard1} as discussed in~\cite{Duari:2026ejq}. The final result is given by
\begin{widetext}
\begin{align}
	2E\frac{dR^{\rm \ hard}}{d^3 p} 
	&= \dfrac{1}{(2\pi)^7}\dfrac{1}{8 E} \int \dfrac{ds}{s} \, dt \int_{\frac{s}{4E} +E}^{\infty} d V  \sum\MB{\mM(s,t)}^2  \int_{E_2^{\rm min}}^{E_2^{\rm max}} ~   \dfrac{dE_2}{\sqrt{(E_2^{\rm max}-E_2) (E_2- E_2^{\rm min})}} ~f_1 f_2 \FB{1\pm f_3}
	\label{Eq_rate_hard3},
\end{align}
\end{widetext}
where we have used the fact that one can write $ \sqrt{ A E_2^2 + B E_2 + C}= s\sqrt{(E_2^{\rm max}-E_2) (E_2- E_2^{\rm min})}$. It should be noted that up to this point the result is exact and none of the integrals can be performed analytically. Using the fact that the photon energy is hard we will replace the distribution functions of the incoming particles by Maxwell-Boltzmann (MB) distribution function.

\subsection*{Pair annihilation process}
In the case of $q \ov{q} \to g \gm $ we have
\begin{align}
	f_1 f_2 &= f_{\text{FD}}^{+} (E_1) \, f_{\text{FD}}^{-} (E_2) \nn \\ &\simeq e^{-(E_1 - \mu)/T} \, e^{-(E_2 + \mu)/T} = e^{-V/T} \nn \\
	f_3 &= f_{\text{BE}}(E_3) = f_{\text{BE}}(V - E) \label{Eq_MB_approx_anni}~,
\end{align}
This shows that effect of $\mu$ disappears for this process after taking MB approximation. In Eq.~\eqref{Eq_rate_hard3} now one can perform the integrations over $E_2$, $V$ and $t$ analytically. To perform $s$-integral one has to further employ the fact that $k_c \ll T$. Thus following previous calculations~\cite{Kapusta:1991qp,Baier:1991em,Haque:2024gva} one can write
\begin{align}
	&2E \frac{dR^{\rm ~anni}}{d^3 p} = \frac{5\alpha\alpha_s}{27\pi^2} \, 2T^2 e^{-E/T} \left[ \ln \frac{4ET}{k_c^2} - 1 - \gamma_E + \frac{\zeta'(2)}{\zeta(2)} \right] \nn \\
	&=-8\dfrac{5 \alpha\alpha_S}{9 \pi^4} { T^2 e^{-E/T}}~\Li \FB{-1} \nn \\
    & ~~~~~~~~~~\times \TB{ \ln \dfrac{4ET}{k_c^2}-1 -\gm_E-\ln 2 + \dfrac{\pLi \FB{-1}}{\Li \FB{-1}}}\label{Eq_rate_anni}
\end{align}
where in the last line we have used the fact that $\pLi (-1) = -\pi^2/12 $. 

\subsection*{Compton scattering}

Compton scattering consists of two subprocesses involving quarks and antiquarks. Let us first consider the  process $q \, g \rightarrow q \, \gamma$. In this case applying MB approximations for the incoming particles we get
\begin{align}
	f_1 f_2 &= f_{\text{FD}}^{+} (E_1) \, f_{\text{BE}}(E_2) \nn \\ &\simeq e^{-(E_1 + E_2 - \mu)/T} = e^{-(V - \mu)/T} \\
	f_3 &= f_{\text{FD}}^{+} (E_3) = f_{\text{FD}}^{+} (V-E)\label{Eq_MB_approx_comp}
\end{align}
Again, from Eq.~\eqref{Eq_rate_hard3}, integration over $E_2$ can be performed analytically yielding a factor of $\pi$. Moreover $V$ and $t$-integrals can also be performed analytically  using Eq.~\eqref{Eq_M2_comp} and  we arrive at the following expression:
\begin{widetext}
\begin{align}
	2E \frac{dR^{q g\rightarrow q \gamma}}{d^3 p} &=  \dfrac{5 \alpha\alpha_S}{9 \pi^4} \dfrac{T e^{-E/T}}{4E} \int_{2 k_c^2}^\infty \dfrac{ds}{s}  \ln \TB{ 1 + \exp \FB{-\dfrac{s - 4\mu E}{4ET}} } \TB{s - 2k_c^2 + 2 s \ln \FB{\dfrac{s-k_c^2}{k_c^2}}}~.
	\label{Eq_sint}
\end{align}
Up to this point, our result is consistent with that of Ref.~\cite{Dumitru:1993us}. In particular, the $s$-integral appearing in the above expression is identical to the one obtained therein. In what follows, we present a consistent treatment of this $s$-integral at finite baryon chemical potential, which, unlike the result of Ref.~\cite{Dumitru:1993us}, yields a hard contribution with a well-behaved $\mu \to 0$ limit. This treatment plays a crucial role in demonstrating of the cutoff independence of the total photon production rate analytically once combined with the known soft contribution given by Eq.~\eqref{Eq_Soft_cont}. To this end, let us write
\begin{align}
	I_s &=  \int_{2 k_c^2}^\infty \dfrac{ds}{s}  \ln \TB{ 1 + \exp \FB{-\dfrac{s - 4\mu E}{4ET}} } \TB{s - 2k_c^2 + 2 s \ln \FB{\dfrac{s-k_c^2}{k_c^2}}} \nn \\
	&= 4 E T \int_{0}^{\infty} \dfrac{dy}{y+ \frac{k_c^2}{2 E T}} \ln \TB{ 1 + e^{-y} e^{-k_c^2/2ET}e^{\mu/T} } \TB{y + 2\FB{y+\frac{k_c^2}{2 E T}} \ln \dfrac{4ET}{k_c^2} + 2 y \ln \FB{y+\frac{k_c^2}{2 E T}}} \nn \\
	&\simeq 4 E T \int_{0}^{\infty} \dfrac{dy}{y} \ln \TB{ 1 + e^{-y} e^{\mu/T} } \TB{y + 2y \ln \dfrac{4ET}{k_c^2} + 2 y \ln y }  \nn \\
	&= 8 E T \sum_{n=1}^{\infty}  \dfrac{(-1)^{n+1} e^{n\mu/T} }{n} \int_{0}^{\infty} dy ~e^{-ny} \TB{\dfrac{1}{2} + \ln \dfrac{4ET}{k_c^2} +  \ln y } \nn \\
	&= 8 E T  \FB{\dfrac{1}{2} + \ln \dfrac{4ET}{k_c^2}  } \sum_{n=1}^{\infty} \dfrac{(-1)^{n+1} e^{n\mu/T} }{n^2} -  8 E T  \sum_{n=1}^{\infty} \dfrac{(-1)^{n+1} e^{n\mu/T} }{n^2}  \FB{-\gm_E  +\ln n } \nn \\
	&=8 E T  \FB{\dfrac{1}{2} -\gm_E + \ln \dfrac{4ET}{k_c^2}  } \sum_{n=1}^{\infty} \dfrac{(-1)^{n+1} e^{n\mu/T} }{n^2} -  8 E T  \sum_{n=1}^{\infty} \dfrac{(-1)^{n+1} e^{n\mu/T} \ln n }{n^2}  \nn \\
	&= - 8 E T ~\Li \FB{-e^{\mu/T}} \TB{ \ln \dfrac{4ET}{k_c^2}+ \dfrac{1}{2} -\gm_E + \dfrac{\pLi \FB{-e^{\mu/T}}}{\Li \FB{-e^{\mu/T}}}}\label{Eq_approx_s_int}
\end{align}
where $\pLi$ is the derivative of polylogarithm with respect to its order. In this derivation in the second line we have used
$y=\dfrac{s- 2k_c^2}{4 E T}$  and in the third line we have used the assumption $k_c \ll T$. Thus we get
\begin{align}
	2E \frac{dR^{q g\rightarrow q \gamma}}{d^3 p} 
	&= -\dfrac{5 \alpha\alpha_S}{9 \pi^4} {2T^2 e^{-E/T}}~\Li \FB{-e^{\mu/T}} \TB{ \ln \dfrac{4ET}{k_c^2}+ \dfrac{1}{2} -\gm_E + \dfrac{\pLi \FB{-e^{\mu/T}}}{\Li \FB{-e^{\mu/T}}}} \label{Eq_quak_comp}
\end{align}
The corresponding contributions from the other Compton processes involving antiquarks can be obtained in an analogous manner and is given by Eq.~\eqref{Eq_quak_comp} with $\mu\to-\mu$. Consequently, the total Compton differential rate for hard photon production is given by
\begin{align}
	2E \frac{dR^{\rm ~comp}}{d^3 p} 
	&= -\dfrac{5 \alpha\alpha_S}{9 \pi^4} {2T^2 e^{-E/T}}~\Li \FB{-e^{\mu/T}} \TB{ \ln \dfrac{4ET}{k_c^2}+ \dfrac{1}{2} -\gm_E + \dfrac{\pLi \FB{-e^{\mu/T}}}{\Li \FB{-e^{\mu/T}}}} \nn \\ 
	&~ ~~~~ -\dfrac{5 \alpha\alpha_S}{9 \pi^4} {2T^2 e^{-E/T}}~\Li \FB{-e^{-\mu/T}} \TB{ \ln \dfrac{4ET}{k_c^2}+ \dfrac{1}{2} -\gm_E + \dfrac{\pLi \FB{-e^{-\mu/T}}}{\Li \FB{-e^{-\mu/T}}}}~.\label{Eq_rate_comp_tot}
\end{align}
Thus the total rate for hard photon production corresponding to physical processes $(2 \to 2)$ in the presence of finite baryon density is obtained by adding Eqs.~\eqref{Eq_rate_anni} and \eqref{Eq_rate_comp_tot} and given by
\begin{align}
	2E \frac{dR^{\rm hard}}{d^3 p} 
	&= ~ - ~8\dfrac{5 \alpha\alpha_S}{9 \pi^4} { T^2 e^{-E/T}}~\Li \FB{-1} \TB{ \ln \dfrac{4ET}{k_c^2}-1 -\gm_E-\ln 2 + \dfrac{\pLi \FB{-1}}{\Li \FB{-1}}} \nn \\
	&~~~~ -\dfrac{5 \alpha\alpha_S}{9 \pi^4} {2T^2 e^{-E/T}}~\Li \FB{-e^{\mu/T}} \TB{ \ln \dfrac{4ET}{k_c^2}+ \dfrac{1}{2} -\gm_E + \dfrac{\pLi \FB{-e^{\mu/T}}}{\Li \FB{-e^{\mu/T}}}} \nn \\ 
	&~~~~ -\dfrac{5 \alpha\alpha_S}{9 \pi^4} {2T^2 e^{-E/T}}~\Li \FB{-e^{-\mu/T}} \TB{ \ln \dfrac{4ET}{k_c^2}+ \dfrac{1}{2} -\gm_E + \dfrac{\pLi \FB{-e^{-\mu/T}}}{\Li \FB{-e^{-\mu/T}}}}~.\label{Eq_hard_ana}
\end{align}
\end{widetext}

\begin{figure}[h]
	\includegraphics[width=0.48\textwidth]{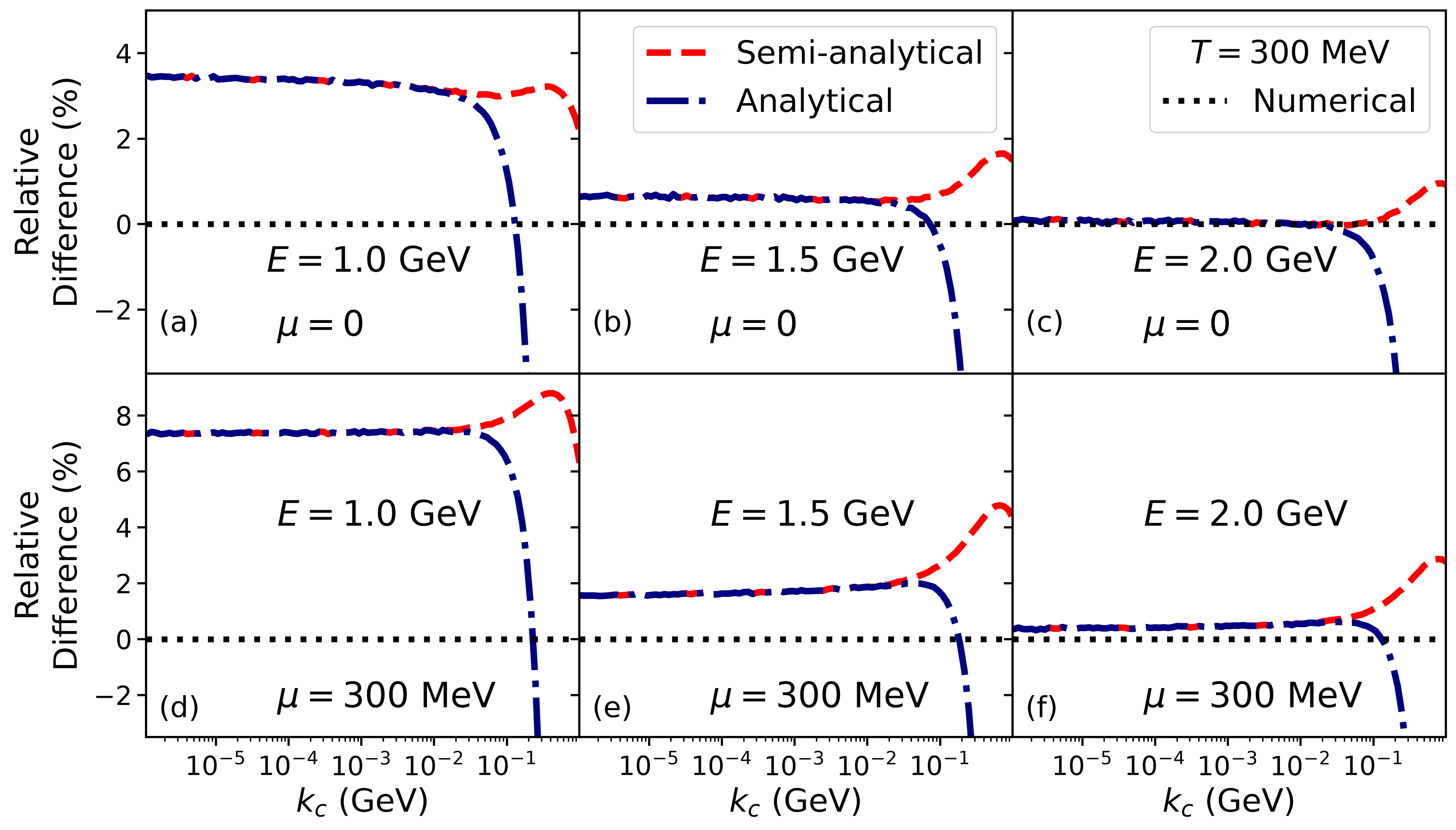}
	\caption{Relative difference of the hard thermal photon rate from its numerical evaluation versus $k_c$ at $T=300$ MeV for different $E$ and $\mu$, with $\alpha_s=0.3$.}
	\label{Fig_hard_rel}
\end{figure}
In Figs.~\ref{Fig_hard_rel}(a)--(f), we plot the relative difference of the hard contribution to the thermal photon rate as a function of $k_c$, comparing the analytical expression of Eq.~\eqref{Eq_hard_ana} and the semi-analytical approach, each evaluated relative to the numerically obtained hard contribution of Eq.~\eqref{Eq_rate_hard2}. By semi-analytical approach we mean that the MB approximations of Eqs.~\eqref{Eq_MB_approx_anni} and \eqref{Eq_MB_approx_comp} are used for the annihilation and Compton processes, respectively. This allows the $E_2$, $V$, and $t$ integrals in Eq.~\eqref{Eq_rate_hard2} to be performed analytically, while the remaining $s$-integral is evaluated numerically without further approximation, as discussed in the derivation leading to Eq.~\eqref{Eq_approx_s_int}. For all the plots we fix the temperature at $T = 300$ MeV. The upper and lower panels correspond to zero chemical potential and $\mu = 300$ MeV, respectively, and the photon energy $E$ increases from left to right. As evident from the plots, our analytical result agrees very well with the semi-analytical result for smaller values of $k_c$, largely independent of the values of $E$, $T$, and $\mu$. For larger $k_c$, however, the analytical result deviates significantly from the semi-analytical result and can even become negative. This behavior is expected because the analytical expression is derived under the assumption $k_c\ll T$, which is well satisfied for the physically relevant values of the cutoff parameter $k_c$ associated with high-energy photon production. Thus, the analytical expression remains reliable within its domain of validity. More importantly, both the MB-approximated results approach the fully numerical result more closely as energy of the photon increases. This is also expected since MB approximations of the incoming participants are  based on the assumption $E\gg T$. This condition is better satisfied for the rightmost plots, resulting in a smaller relative difference. The presence of a finite chemical potential further weakens the MB approximation at fixed $E$ and $T$, which explains the increase in the relative differences observed in the lower panels.

\section{Total photon rate from hot and dense QGP}

Adding the analytical hard contribution obtained in Eq.~\eqref{Eq_hard_ana} with the soft contribution given by Eq.~\eqref{Eq_Soft_cont} and using the expression for the thermal quark mass expressed in terms of $\Li$-function as given in Eq.~\eqref{Eq_ML}, we finally obtain a cutoff independent analytical expression for thermal photon production rate from a QGP medium with finite quark chemical potential, i.e.
\begin{widetext}
\begin{align}
	2E \frac{dR^{\rm total}}{d^3 p} 
	&= ~ - ~8\dfrac{5 \alpha\alpha_S}{9 \pi^4} {T^2 e^{-E/T}}~\Li \FB{-1} \TB{  \ln \dfrac{2ET}{M^2} -1 -\gm_E-\ln 2 + \dfrac{\pLi \FB{-1}}{\Li \FB{-1}}} \nn \\
	&~~~~ -\dfrac{5 \alpha\alpha_S}{9 \pi^4} {2T^2 e^{-E/T}}~\Li \FB{-e^{\mu/T}} \TB{ \ln \dfrac{2ET}{M^2}+ \dfrac{1}{2} -\gm_E + \dfrac{\pLi \FB{-e^{\mu/T}}}{\Li \FB{-e^{\mu/T}}}} \nn \\ 
	&~~~~ -\dfrac{5 \alpha\alpha_S}{9 \pi^4} {2T^2 e^{-E/T}}~\Li \FB{-e^{-\mu/T}} \TB{ \ln \dfrac{2ET}{M^2}+ \dfrac{1}{2} -\gm_E + \dfrac{\pLi \FB{-e^{-\mu/T}}}{\Li \FB{-e^{-\mu/T}}} }~.\label{Eq_rate_mu_1}
\end{align}
Eq.~\eqref{Eq_rate_mu_1} can be expressed in a more familiar fashion assuming small values of $\mu$. In this limit we can write $\dfrac{\pLi \FB{-e^{\pm\mu/T}}}{\Li \FB{-e^{\pm\mu/T}}} \approx \dfrac{\pLi \FB{-1}}{\Li \FB{-1}}  $ and then use the fact that $\Li (-e^x)+\Li (-e^{-x}) = -\pi^2/6 - x^2/2$. Thus we arrive at
\begin{align}
	2E \frac{dR^{\rm total}}{d^3 p} 
	&\simeq \dfrac{5 \alpha\alpha_S}{9 \pi^2} {T^2 e^{-E/T}} \TB{ \ln \dfrac{2ET}{M^2}-\half -\gm_E-\dfrac{2}{3}\ln 2 + \dfrac{\pLi \FB{-1}}{\Li \FB{-1}}}  + \dfrac{5 \alpha\alpha_S}{9 \pi^4} {\mu^2 e^{-E/T}}  \TB{ \ln \dfrac{2ET}{M^2}+ \dfrac{1}{2} -\gm_E + \dfrac{\pLi \FB{-1}}{\Li \FB{-1}}} \nn \\
	&=  \dfrac{5 \alpha\alpha_S}{9 \pi^2} {T^2 e^{-E/T}} \TB{ \ln \dfrac{2ET}{M^2}-\half -\gm_E+\dfrac{1}{3}\ln 2 + \dfrac{\zeta^\prime(2)}{\zeta(2)}}  + \dfrac{5 \alpha\alpha_S}{9 \pi^4} {\mu^2 e^{-E/T}}  \TB{ \ln \dfrac{2ET}{M^2}+ \dfrac{1}{2} -\gm_E + \ln 2 + \dfrac{\zeta^\prime(2)}{\zeta(2)}} \nn \\
		&=  \dfrac{5 \alpha\alpha_S}{9 \pi^2} {T^2 e^{-E/T}} \TB{ \ln \dfrac{2.91182~ET}{g^2\FB{T^2 + \frac{\mu^2}{\pi^2}} }  }  + \dfrac{5 \alpha\alpha_S}{9 \pi^2} \dfrac{\mu^2}{\pi^2}{ e^{-E/T}}  \TB{ \ln \dfrac{12.5645~ET}{g^2\FB{T^2 + \frac{\mu^2}{\pi^2}} }}~.	\label{Eq_rate_mu_2}
\end{align}
\end{widetext}
In the above expression, the first term matches with the thermal photon rate obtained in~\cite{Kapusta:1991qp,Baier:1991em,Haque:2024gva} when one puts $\mu=0$. The second term is the additional contribution coming from the presence of finite baryon chemical potential. It is also evident that Eq.~\eqref{Eq_rate_mu_2} is quite different from  the parametrized formula given in Ref.~\cite{Traxler:1994hy}.

\begin{figure}[h]
    \includegraphics[width=0.45\textwidth]{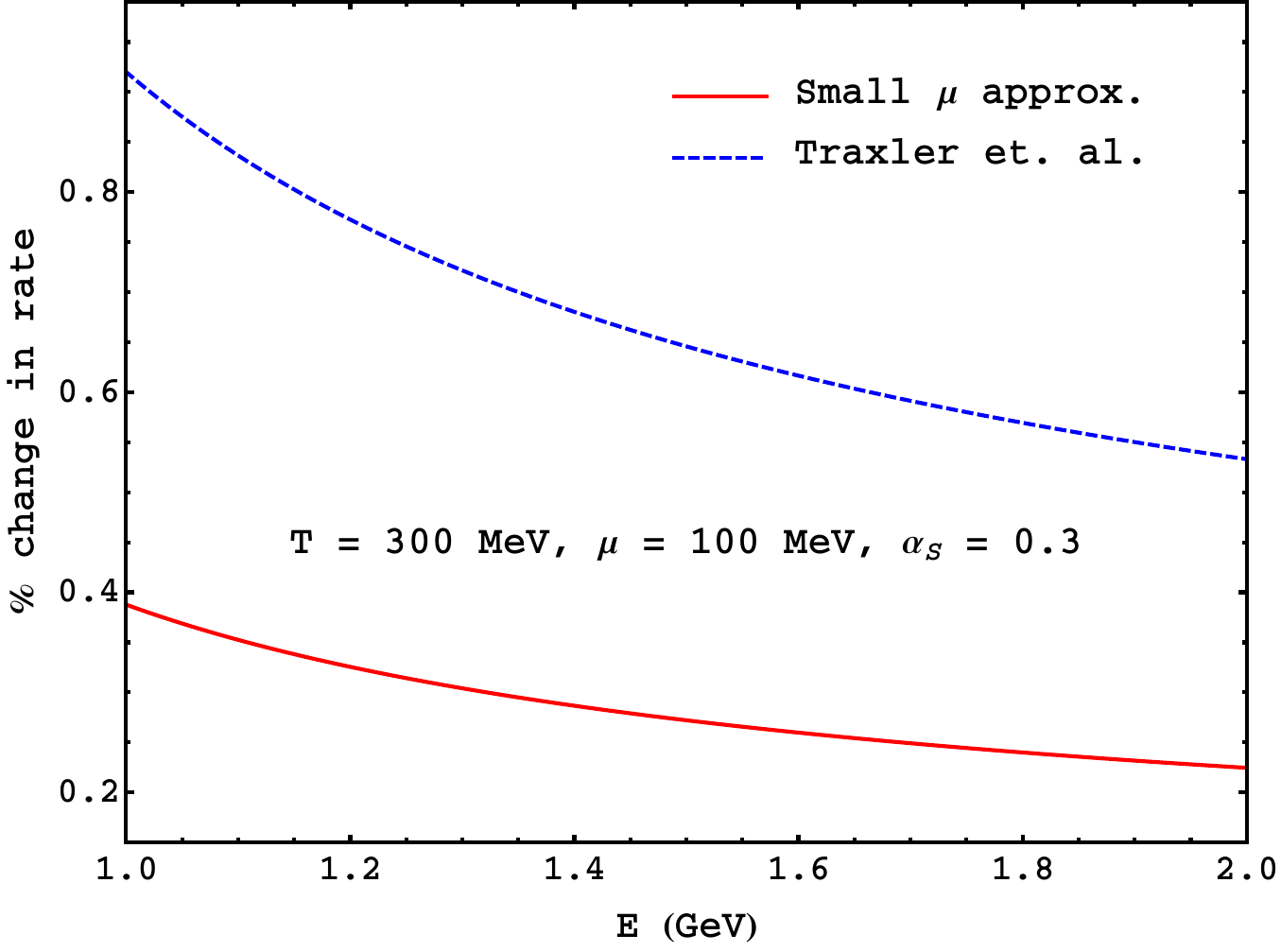}
	\caption{Percentage deviation of the total thermal photon production rate obtained using the phenomenological parametrization of Traxler \textit{et al.}~\cite{Traxler:1994hy} relative to the result obtained in this work, for $T=300$ MeV and $\mu=300$ MeV.
}
	\label{Fig_percentage}
\end{figure}
In Fig.~\ref{Fig_percentage}, we display the percentage deviation of the total thermal photon production rate obtained using the approximate expression given in Eq.~\eqref{Eq_rate_mu_2} and the parametrization of Ref.~\cite{Traxler:1994hy}, relative to the total rate given by Eq.~\eqref{Eq_rate_mu_1} for $T=300$ MeV and $\mu=100$ MeV.  It is evident that Eq.~\eqref{Eq_rate_mu_2} provides a systematically better agreement with the exact result than the parametrization of Ref.~\cite{Traxler:1994hy}. The deviation of the latter from the total photon rate  is particularly pronounced at lower photon energies.  Moreover, we have checked that the efficiency of the parametrization of Ref.~\cite{Traxler:1994hy} worsens considerably with increasing $\mu$ reaching up to 7\% (depending on the photon energy) once $\mu$ becomes comparable to $T$.

\begin{figure}[h]
	\includegraphics[width=0.45\textwidth]{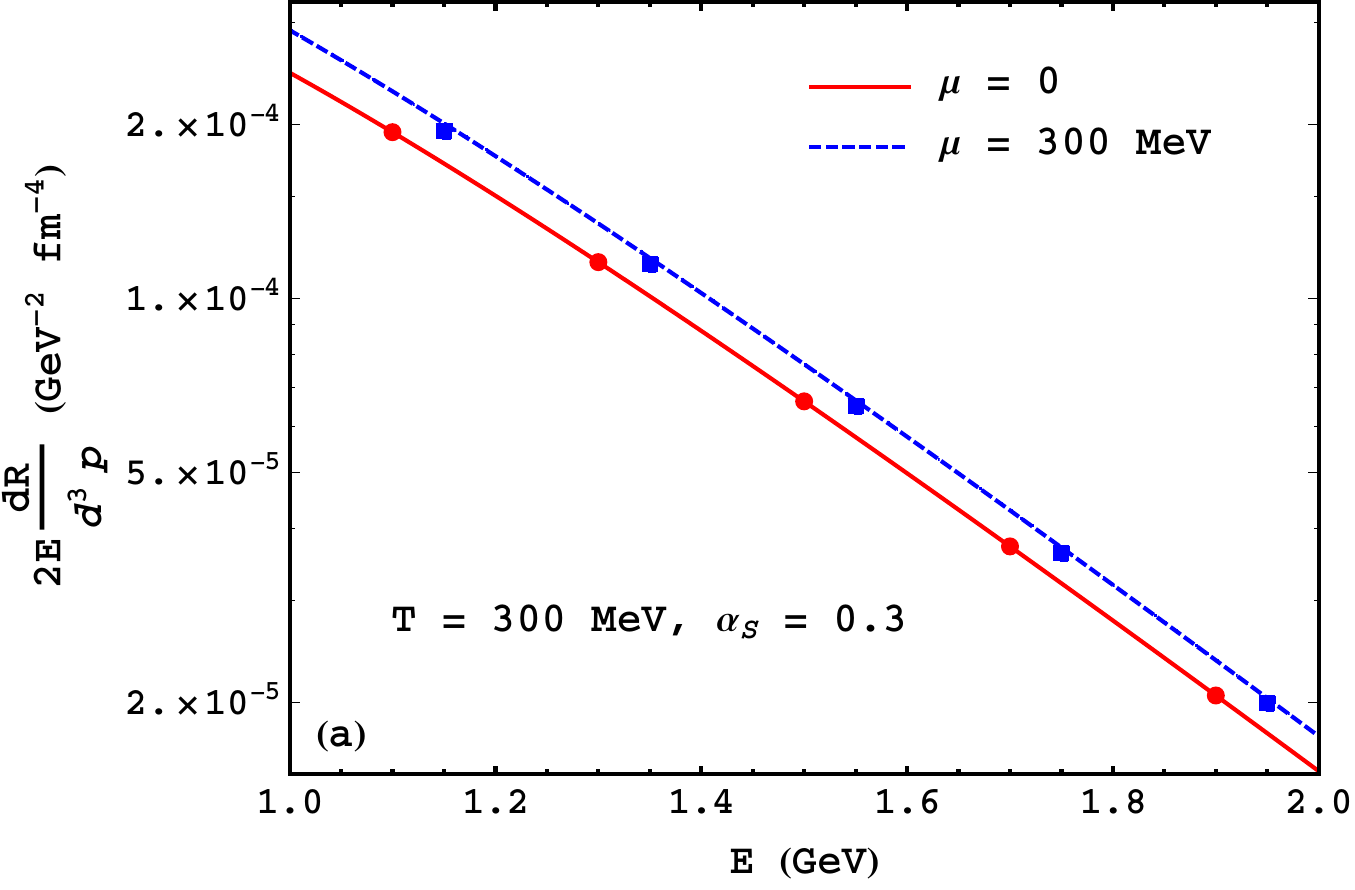} ~~~~
	\includegraphics[width=0.45\textwidth]{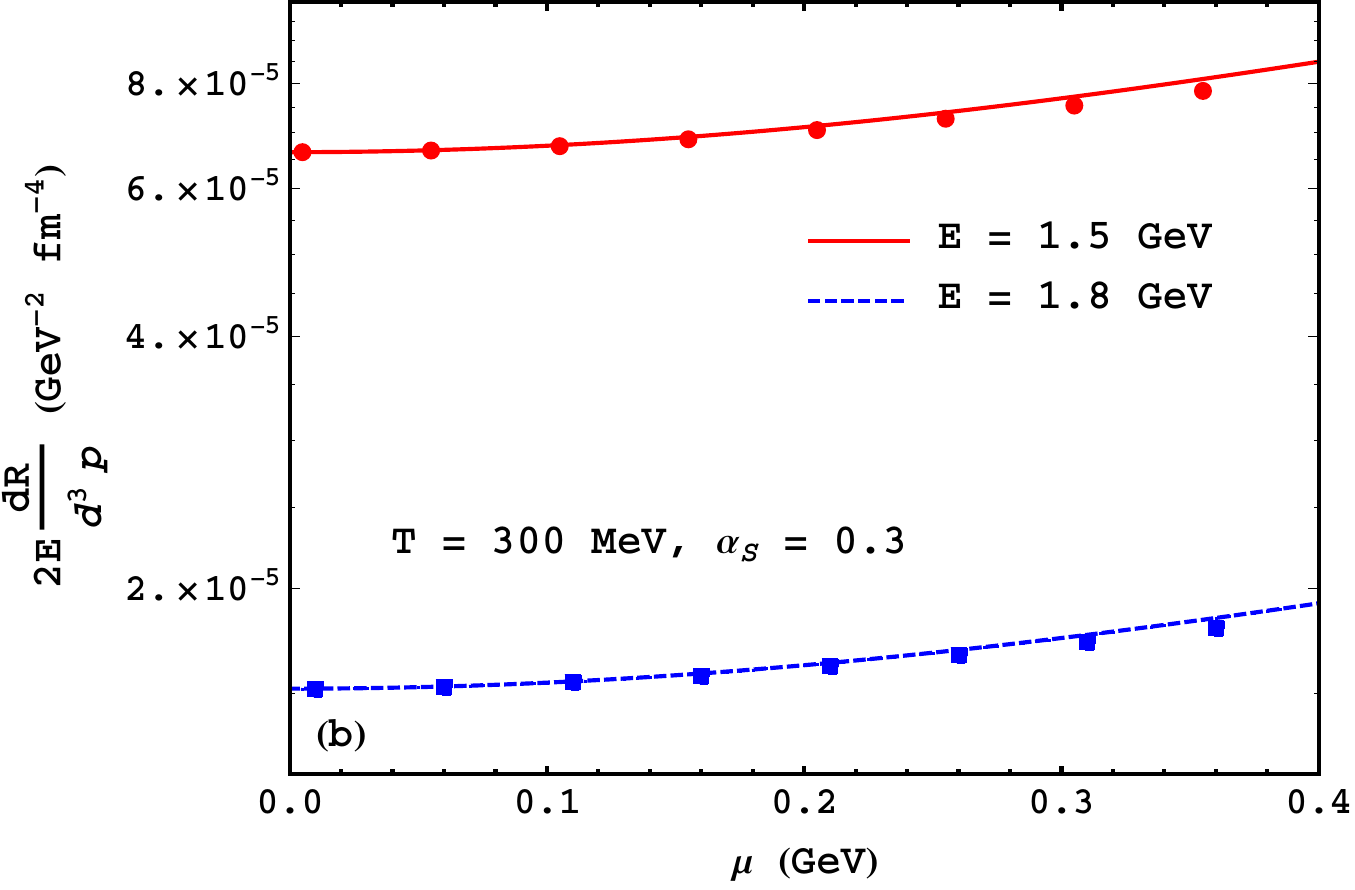}
	\caption{Thermal photon production rate at $T = 300$ MeV (a) as a function of photon energy for different values baryon density (b) as a function of \(\mu\) for different photon energies. The results obtained using small $\mu$ approximation as given by Eq.~\eqref{Eq_rate_mu_2} are shown by points. All results are obtained for $\alpha_S = 0.3$.}
	\label{Fig_Rate}
\end{figure}

Figs.~\ref{Fig_Rate}(a) and (b) illustrate the total production rate of high-energy photons at $T = 300~\mathrm{MeV}$ as a function of photon energy for various baryon densities, and as a function of the chemical potential $\mu$  for particular values of photon energy, respectively. The photon spectra seen in Fig.~\ref{Fig_Rate} (a) exhibit an exponential suppression with increasing energy as expected in a thermal system. The presence of a finite baryon density yields a notable enhancement in the emission rate. This behavior stems from modifications of the underlying quark and antiquark distribution functions induced by a non-zero $\mu$ which alters the accessible phase space and scattering probabilities governing photon production. Consequently, a finite chemical potential increases the effective density of active participants in the plasma, resulting in an enhanced net photon yield. Furthermore, Fig.~\ref{Fig_Rate}(b) demonstrates that the production rate monotonically increases with $\mu$, with the enhancement being particularly pronounced at lower photon energies.

\section{Summary \& outlook}
\label{Sec_summary}

In this article, we have derived, for the first time, an analytic proof of the cancellation of the intermediate momentum cutoff $k_c$ between the hard and soft contributions to the thermal photon production rate from a QGP at finite quark chemical potential $\mu$. This cutoff independence was previously established only numerically in Refs.~\cite{Traxler:1994hy,Duari:2026ejq}. We have shown that the hard contribution evaluated by approximating the distribution functions of both incoming participants in the pair-annihilation and QCD Compton processes by Maxwell--Boltzmann forms, valid in the hard-photon kinematic regime $E_1+E_2>E\gg T$.  Combining this with the known soft contribution  yields a closed-form, cutoff-independent expression for the total photon rate, as given by Eq.~\eqref{Eq_rate_mu_1}. A compact small-$\mu$ expansion given by Eq.~\eqref{Eq_rate_mu_2} reproduces the $\mu = 0$ result of Refs.~\cite{Kapusta:1991qp,Baier:1991em} and isolates a new $\mathcal{O}(\mu^2)$ term arising purely from finite baryon density. We find that this analytic result agrees markedly well with the full numerical rate than the widely used phenomenological parametrization of Traxler \textit{et al.}~\cite{Traxler:1994hy}. The improvement is most pronounced at low photon energies and moderate $\mu/T$, precisely the regime relevant to the baryon-rich matter probed in RHIC beam-energy-scan and upcoming FAIR/NICA experiments. Physically, a finite quark chemical potential enhances the photon yield by increasing the effective phase-space density of quarks and antiquarks available for annihilation and Compton scattering.  It has also been shown that for a particular temperature the rate monotonically increases as a function of quark chemical potential $\mu$.

Since Eq.~\eqref{Eq_rate_mu_1} is given in a closed analytic form, it can be directly incorporated in relativistic hydrodynamic simulations of heavy-ion collisions without the computational overhead of the fully numerical rate or the systematic bias of the older parametrization. This makes our result immediately useful for extraction of direct-photon spectra and flow observables at finite $\mu$ using space-time evolution of the QGP medium. Natural extensions of this work include incorporating the effects of a chirally imbalanced medium~\cite{Duari:2026ejq}, next-to-leading-order QCD corrections, and a consistent treatment of the photon rate across the full QCD phase diagram.

\bibliography{reference} 

	\end{document}